\documentstyle[epsf]{l-aa}

\newcommand{\mincir}{\ \raise -2.truept\hbox{\rlap{\hbox{$\sim$}}\raise5.truept
	\hbox{$<$}\ }}			
\newcommand{\magcir}{\ \raise -2.truept\hbox{\rlap{\hbox{$\sim$}}\raise5.truept
 	\hbox{$>$}\ }}		
\twocolumn
\begin{document}

   \thesaurus{03         % A&A Section 3: extragalactic Astronomy
              (11.17.3)  % quasars: general
             }

   \title{ The optical variability of QSOs. II. The wavelength dependence
   \thanks{Based on material collected with the UKSTU,
    the ESO-La Silla and the Mount Palomar telescopes and on COSMOS
scans.
Table 1 is only available in electronic form at the CDS via anonymous
ftp to cdsarc.u-strasbg.fr (130.79.128.5) or via
http://cdsweb.u-strasbg.fr/Abstract.html } 
	}

   \author{S. Cristiani\inst{1}
	\and S. Trentini\inst{1}
	\and F. La Franca\inst{2}
	\and P. Andreani\inst{1}
          }
 
   \offprints{S. Cristiani}

   \institute{Dipartimento di Astronomia, Universit\`{a} di Padova,
	    Vicolo dell'Osservatorio 5,\\I-35122 Padova, Italy
	\and 
	Dipartimento di Fisica, Universit\`a degli 
	Studi ``Roma Tre'', Via della Vasca Navale 84, I-00146 Roma, Italy}
 
   \date{Received 25 April 1996; accepted \dots}
 
   \maketitle
 
   \begin{abstract}

The long-term variability of a sample of 149 optically selected QSOs 
in the field of the Selected Area 94 
has been studied in the R-band.
The relations between variability and luminosity and between variability and
redshift have been investigated by means of ``robust'' statistical estimators,
allowing to disentangle the effects of the measurement errors. 
The results are compared with the corresponding properties of the variability
in the B-band for the same sample. 
An anti-correlation between the R-band variability and the intrinsic 
luminosity is found, 
analogously to what is observed in the B-band.
The amplitude of the R-band variability
turns out to be smaller (of a factor $1.13 \pm 0.05$) 
than the B-band variability.
The implications in terms of the black-hole, starburst and microlensing 
models are discussed.


   \end{abstract}
 
\section{Introduction}

The detailed study of statistically well-defined samples of optically selected
QSOs is beginning to pin down their ensemble variability properties. 
An anti-correlation between variability and luminosity
has been observed, such that more luminous QSOs show a smaller
amplitude of the variations (\cite{hook:94}, Cristiani et al. 1996).
It starts also to emerge from the statistical noise a positive correlation
between the amplitude of the variability and the redshift 
(\cite{gtv:91}, \cite{dario:94}, Cristiani et al. 1996, \cite{cid:96}).
This latter dependence may be interpreted as an anti-correlation between
amplitude of the variability and wavelength in the individual objects
(Cristiani et al. 1996, 
\cite{cid:96}, \cite{dicl:96}), as suggested by observations of
relatively small samples of objects (\cite{per:82}, Edelson et al.
\cite{ede:kro},
Kinney et al. \cite{kin:91}, Paltani \& Courvoisier \cite{pal:cou}).
Both properties bear important implications for the understanding of the AGN
energy generation mechanism and distinguishing among the competing variability
models based on the black-hole (\cite{Rees:84}), starburst (\cite{ter:tt}) and
microlensing (\cite{hawk:93}, \cite{hawk:96}) scenarios.
At the same time they are central for the quantification of the biases affecting
samples selected on the sole basis of variability (\cite{verhaw:95}) or on the
basis of multicolor data, if the photometry was carried out at different epochs
in the various wavebands (e.g. \cite{Warr:91}). 

However the variability-redshift correlation could be due to an evolutionary
effect, in the sense of a higher activity in the early phases of the AGN life.
The only way to remove such an uncertainty is to check directly the variability
properties of individual objects, by observing them simultaneously in two or
more wavebands. 
For this reason we have investigated the R-band variability of the SA94 QSO
sample, the same that in Cristiani et al. 1996 (in the following \cite{papI})
had been studied in the B band. 

We report in Section 2 the description of the SA94 QSO sample and the
photometric material used to investigate its variability; in Section 3 the
ensemble structure function is introduced as a statistical measure of the
variability properties and the detected trends are further analysed with the
aid of other statistical indices; the results and implications are discussed in
Section 4. 

Throughout the paper we have assumed cosmological constants
$H_{o} = 50$ Km sec$^{-1}$ Mpc$^{-1}$ , $ q_{o} = 0.5 $.

\section{The quasars sample}

\subsection {Definition of the sample}

The SA94 sample is made up of 149 optically selected quasars \footnote{In the
present paper quasars are defined as objects with a starlike nucleus, broad
emission lines, brighter than $M_B = -23$ mag, applying the K-corrections
computed on the basis of the composite spectrum of Cristiani and Vio (1990),
and dereddened for galactic extinction according to Burstein and Heiles
(1982).} listed in Table~\ref{tab:qsos}, observed in a rectangular area of 
the sky within the limits $2^{h} 43^{m} 51.2^{s} < \alpha < 
2^{h} 59^{m} 14.3^{s}$, $-2^{\circ} 03' 23.8'' < \delta <
2^{\circ} 32' 11.0'' $ (Epoch 1950.0), covering 17.66 sq. deg.

They have been mainly discovered in the UVx and objective-prism surveys 
described in \cite{papI} and in the references to Table~\ref{tab:qsos}.

\begin{table}
\caption[ ]{List of the QSOs in the SA94}  \label{tab:qsos}
\begin{flushleft}
\begin{tabular}[ ]{cccccc}
\hline
\bf $\alpha $ & \bf $\delta $ & \bf z & $R'$  & $IDX$& \bf Ref.\\
\hline
 2 43 54.36 &  $+$1 27 42.9 & 1.904  &   19.03  &   $-$0.02 & e \\
 2 43 55.42 &  $-$0 57 30.5 & 2.103  &   19.55  &   $ $ --  & e \\
 2 43 58.80 &  $-$0 07 03.9 & 1.305  &   18.53  &   $+$0.02 & a \\
 2 43 59.21 &  $-$0 44 47.3 & 2.147  &   18.55  &   $+$0.20 & a \\
 2 43 59.35 &  $-$0 47 12.2 & 1.726  &   19.21  &   $+$0.07 & e \\
 2 44 01.14 &  $+$1 16 08.9 & 2.032  &   18.74  &   $-$0.02 & a \\
 2 44 02.07 &  $-$0 21 23.3 & 1.815  &   18.16  &   $+$0.02 & a \\
 2 44 04.84 &  $-$0 53 38.2 & 0.859  &   19.72  &   $-$0.02 & e \\
 2 44 06.25 &  $-$0 15 09.7 & 2.315  &   19.56  &   $ $ --  & e \\
 2 44 10.58 &  $-$1 44 22.1 & 0.506  &   18.50  &   $-$0.02 & e \\
%%  2 44 12.19 &  $-$2 02 22.3 & 1.552  &   19.76  &   $ $     & e \\
 2 44 14.86 &  $-$1 58 07.0 & 1.784  &   17.62  &   $+$0.04 & a \\
 2 44 17.87 &  $-$0 57 29.3 & 2.172  &   19.02  &   $+$0.63 & e \\
 2 44 19.04 &  $-$1 12 03.0 & 0.467  &   16.56  &   $-$0.01 & c \\
 2 44 22.41 &  $+$1 46 40.2 & 1.945  &   18.80  &   $+$0.03 & a \\
 2 44 27.45 &  $-$0 09 01.2 & 2.137  &   18.98  &   $-$0.02 & e \\
 2 45 06.14 &  $-$1 04 50.9 & 2.125  &   19.46  &   $-$0.04 & e \\
 2 45 14.58 &  $-$1 00 39.2 & 1.918  &   19.51  &   $ $ --  & e \\
 2 45 15.22 &  $-$0 14 16.8 & 1.859  &   19.49  &   $+$0.06 & e \\
 2 45 16.43 &  $+$1 19 20.4 & 2.310  &   18.46  &   $-$0.02 & c \\
 2 45 18.18 &  $-$1 52 47.8 & 1.474  &   19.73  &   $ $ --  & e \\
 2 45 20.98 &  $-$1 44 54.0 & 1.937  &   18.47  &   $+$0.00 & a \\
 2 45 22.87 &  $-$0 28 24.9 & 2.118  &   18.02  &   $+$0.00 & a \\
 2 45 27.96 &  $-$0 52 44.3 & 0.812  &   19.19  &   $+$0.05 & e \\
 2 45 33.58 &  $+$0 23 23.9 & 0.835  &   19.13  &   $-$0.01 & e \\
 2 45 47.58 &  $-$0 38 14.4 & 1.450  &   19.16  &   $+$0.00 & e \\
 2 45 49.38 &  $+$0 23 25.5 & 1.015  &   19.56  &   $-$0.04 & e \\
 2 45 57.55 &  $+$0 37 06.0 & 1.598  &   19.10  &   $+$0.01 & e \\
 2 46 00.83 &  $-$0 58 43.3 & 1.822  &   18.07  &   $+$0.01 & a \\
 2 46 05.12 &  $+$2 11 16.9 & 1.267  &   17.90  &   $+$0.01 & e \\
 2 46 07.49 &  $-$0 24 55.4 & 1.684  &   18.43  &   $+$0.01 & e \\
 2 46 07.59 &  $-$0 48 15.0 & 2.239  &   18.24  &   $+$0.02 & e \\
 2 46 21.63 &  $+$0 10 40.8 & 1.017  &   19.27  &   $+$0.36 & e \\
 2 46 23.61 &  $-$1 08 30.1 & 1.709  &   18.61  &   $-$0.02 & e \\
 2 46 33.53 &  $-$1 46 34.1 & 1.152  &   18.69  &   $+$0.13 & e \\
 2 46 33.65 &  $-$0 19 14.8 & 2.249  &   19.59  &   $ $ --  & e \\
 2 46 47.09 &  $+$1 56 38.4 & 1.953  &   18.68  &   $-$0.02 & a \\
 2 46 50.66 &  $-$1 55 39.2 & 1.434  &   18.88  &   $-$0.02 & e \\
 2 46 52.63 &  $-$0 32 13.1 & 2.475  &   19.26  &   $-$0.03 & e \\
 2 46 54.28 &  $+$0 57 00.4 & 0.954  &   18.34  &   $+$0.20 & a \\
 2 46 55.79 &  $-$0 33 28.5 & 1.419  &   18.24  &   $+$0.01 & e \\
 2 46 59.51 &  $-$1 47 01.9 & 2.337  &   19.61  &   $ $ --  & e \\
 2 47 09.68 &  $+$1 41 16.8 & 2.690  &   18.54  &   $+$0.01 & a \\
 2 47 09.99 &  $+$0 20 54.1 & 1.480  &   19.54  &   $ $ --  & e \\
 2 47 10.52 &  $+$1 10 27.3 & 1.032  &   18.87  &   $-$0.02 & d \\
 2 47 20.03 &  $+$0 49 25.1 & 0.584  &   18.78  &   $+$0.09 & d \\
 2 47 39.76 &  $+$1 29 41.1 & 2.054  &   18.83  &   $-$0.03 & d \\
 2 47 57.22 &  $-$0 20 23.1 & 1.458  &   17.69  &   $+$0.05 & d \\
 2 48 03.30 &  $+$1 05 49.8 & 1.828  &   19.46  &   $-$0.01 & d \\
 2 48 05.66 &  $-$0 59 59.9 & 1.845  &   18.18  &   $-$0.01 & d \\
 2 48 06.79 &  $-$1 00 10.3 & 2.422  &   19.54  &   $ $ --  & e \\
 2 48 14.96 &  $-$0 10 12.8 & 0.766  &   18.35  &   $+$0.02 & d \\
 2 48 23.20 &  $+$0 54 35.0 & 1.708  &   18.83  &   $+$0.23 & d \\
 2 48 26.70 &  $+$0 35 43.5 & 0.828  &   19.15  &   $+$0.12 & d \\
 2 48 28.94 &  $-$0 06 18.9 & 1.435  &   18.48  &   $+$0.03 & d \\
 2 48 34.91 &  $+$1 30 39.3 & 0.815  &   19.36  &   $+$0.03 & d \\
 2 48 55.43 &  $-$0 39 09.1 & 2.329  &   19.42  &   $+$0.28 & d \\
 2 49 12.07 &  $-$0 58 57.3 & 1.383  &   19.10  &   $+$0.06 & d \\
\hline
\end{tabular}
\end{flushleft}
\end{table}

\setcounter{table}{0}
\begin{table}
\caption[ ]{List of the QSOs in the SA94 - end}  
\begin{flushleft}
\begin{tabular}[ ]{cccccc}
\hline
\bf $\alpha $ & \bf $\delta $ & \bf z & $R'$  & $IDX$ & \bf Ref.\\
\hline
 2 49 13.67 &  $-$0 52 52.9 & 0.817  &   19.53  &   $+$0.11 & d \\
 2 49 15.52 &  $-$0 58 55.1 & 1.569  &   18.75  &   $+$0.02 & d \\
%% 2 49 16.86 &  $+$0 45 22.1 & 1.824  &   19.78  &   $ $     & d \\
 2 49 17.04 &  $+$1 18 40.2 & 2.981  &   18.99  &   $ $ --  & e \\
 2 49 21.88 &  $+$0 44 49.6 & 0.470  &   18.11  &   $+$0.11 & d \\
 2 49 36.07 &  $-$0 06 27.3 & 2.099  &   19.38  &   $-$0.02 & d \\
 2 49 42.41 &  $+$2 22 57.2 & 2.805  &   17.94  &   $+$0.15 & a \\
 2 49 46.50 &  $+$0 15 19.3 & 1.678  &   19.30  &   $+$0.22 & d \\
 2 49 47.27 &  $-$0 06 16.5 & 0.810  &   16.97  &   $+$0.03 & d \\
 2 49 54.48 &  $+$0 18 53.6 & 1.106  &   18.48  &   $-$0.01 & d \\
 2 49 55.30 &  $+$0 48 33.7 & 2.010  &   19.23  &   $+$0.00 & d \\
 2 50 04.74 &  $-$0 58 42.9 & 1.007  &   19.13  &   $-$0.01 & d \\
 2 50 04.98 &  $-$1 06 21.5 & 0.846  &   19.19  &   $+$0.10 & d \\
 2 50 13.01 &  $+$1 40 49.4 & 2.637  &   18.14  &   $+$0.19 & d \\
 2 50 24.34 &  $-$1 14 34.8 & 1.251  &   18.65  &   $+$0.02 & d \\
 2 50 34.48 &  $+$1 08 19.5 & 1.331  &   19.55  &   $ $ --  & d \\
 2 50 40.67 &  $+$2 03 21.0 & 1.393  &   18.16  &   $-$0.01 & b \\
 2 50 40.86 &  $-$0 51 15.6 & 0.889  &   18.62  &   $+$0.30 & d \\
 2 50 41.90 &  $+$0 55 47.0 & 1.030  &   18.58  &   $+$0.02 & d \\
 2 50 47.21 &  $-$1 46 26.6 & 0.673  &   18.80  &   $-$0.01 & e \\
 2 50 49.92 &  $-$0 09 42.1 & 1.214  &   18.98  &   $+$0.00 & d \\
 2 50 51.40 &  $-$0 39 08.3 & 1.363  &   18.93  &   $+$0.02 & d \\
 2 50 54.00 &  $-$1 46 51.3 & 2.550  &   18.62  &   $-$0.01 & e \\
 2 50 54.55 &  $+$1 54 32.2 & 1.925  &   18.98  &   $+$0.04 & a \\
 2 50 58.05 &  $+$0 04 12.6 & 1.810  &   18.81  &   $+$0.02 & d \\
 2 51 07.14 &  $-$0 01 01.7 & 1.677  &   17.99  &   $+$0.07 & d \\
 2 51 22.28 &  $-$0 01 13.5 & 1.688  &   19.22  &   $-$0.01 & d \\
 2 51 23.31 &  $-$0 23 34.7 & 0.757  &   19.02  &   $+$0.03 & d \\
 2 51 27.40 &  $+$0 17 05.5 & 1.986  &   19.30  &   $-$0.04 & d \\
 2 51 49.09 &  $-$0 04 10.6 & 1.213  &   19.25  &   $+$0.03 & d \\
 2 51 53.15 &  $-$1 53 34.9 & 1.422  &   19.18  &   $+$0.27 & e \\
 2 51 59.35 &  $-$1 01 42.1 & 1.955  &   19.67  &   $ $ --  & d \\
 2 51 59.44 &  $-$0 54 29.1 & 0.433  &   17.56  &   $+$0.03 & d \\
 2 52 08.14 &  $+$1 41 09.8 & 0.620  &   17.45  &   $+$0.01 & d \\
 2 52 31.66 &  $+$0 13 15.5 & 0.354  &   17.37  &   $+$0.20 & d \\
 2 52 39.29 &  $-$0 05 27.3 & 1.885  &   19.30  &   $+$0.18 & d \\
 2 52 40.10 &  $+$1 36 21.8 & 2.457  &   17.40  &   $+$0.02 & d \\
 2 52 55.32 &  $-$0 14 25.5 & 1.426  &   19.25  &   $+$0.00 & d \\
 2 53 12.89 &  $+$0 26 09.5 & 0.921  &   18.32  &   $+$0.04 & d \\
 2 53 25.53 &  $+$0 41 06.9 & 0.847  &   18.24  &   $+$0.06 & d \\
 2 53 27.93 &  $-$1 27 48.2 & 1.260  &   18.01  &   $+$0.00 & e \\
 2 53 28.19 &  $+$0 40 50.8 & 0.531  &   18.93  &   $+$0.58 & d \\
 2 53 32.65 &  $+$0 58 34.3 & 1.347  &   18.19  &   $+$0.07 & d \\
 2 53 34.59 &  $+$1 44 30.0 & 1.439  &   18.83  &   $+$0.00 & d \\
%% 2 53 39.27 &  $+$0 03 04.2 & 2.012  &   19.75  &   $ $     & e \\
 2 53 39.85 &  $+$0 27 37.1 & 0.916  &   18.38  &   $+$0.02 & d \\
 2 53 44.04 &  $-$1 38 42.2 & 0.878  &   16.49  &   $+$0.01 & a \\
 2 53 45.96 &  $-$0 57 05.1 & 0.720  &   18.44  &   $+$1.06 & e \\
 2 54 07.94 &  $-$1 37 48.1 & 2.684  &   18.89  &   $+$0.06 & a \\
 2 54 10.86 &  $+$0 00 43.5 & 2.242  &   17.82  &   $+$0.01 & d \\
 2 54 24.21 &  $+$0 42 45.8 & 1.115  &   19.04  &   $-$0.01 & d \\
 2 54 26.14 &  $+$1 26 12.6 & 1.793  &   18.99  &   $+$0.10 & d \\
 2 54 29.28 &  $-$1 14 18.8 & 0.876  &   18.97  &   $-$0.03 & d \\
 2 54 32.36 &  $-$0 22 54.8 & 1.585  &   18.99  &   $-$0.02 & d \\
 2 54 40.26 &  $-$1 13 58.7 & 1.866  &   19.00  &   $+$0.00 & d \\
 2 54 40.98 &  $+$1 13 39.0 & 1.089  &   19.17  &   $+$0.00 & d \\
 2 54 43.83 &  $-$0 57 36.9 & 1.032  &   18.86  &   $+$0.46 & d \\
 2 54 51.41 &  $-$0 10 46.2 & 1.250  &   19.03  &   $+$0.20 & d \\
 2 54 53.37 &  $+$0 03 45.5 & 1.601  &   19.57  &   $ $ --  & c \\
 2 55 13.65 &  $-$1 31 46.7 & 1.520  &   17.73  &   $+$0.02 & d \\
 2 55 17.59 &  $+$0 08 46.6 & 1.498  &   18.68  &   $+$0.04 & d \\
 2 55 28.50 &  $+$1 52 05.2 & 1.623  &   18.76  &   $+$0.04 & d \\
\hline
\end{tabular}
\end{flushleft}
\end{table}
\begin{table}
\begin{flushleft}
\begin{tabular}[t]{cccccc}
\hline
\bf $\alpha $ & \bf $\delta $ & \bf z & $R'$  & $IDX$ & \bf Ref.\\
\hline
 2 55 30.75 &  $-$0 22 58.4 & 1.557  &   19.24  &   $+$0.25 & d \\
 2 55 41.94 &  $-$0 15 32.0 & 1.318  &   18.66  &   $+$0.04 & d \\
 2 55 45.79 &  $-$0 20 04.0 & 2.094  &   19.58  &   $ $ --  & d \\
 2 56 11.30 &  $-$1 07 08.9 & 0.905  &   18.59  &   $+$0.23 & d \\
 2 56 12.61 &  $+$1 50 08.6 & 0.706  &   19.03  &   $-$0.01 & d \\
 2 56 14.72 &  $+$1 40 28.8 & 0.608  &   18.36  &   $-$0.01 & d \\
%% 2 56 20.64 &  $+$0 30 25.9 & 1.569  &   19.79  &   $ $     & d \\
 2 56 31.80 &  $-$0 00 33.3 & 3.367  &   17.17  &   $+$0.00 & a \\
 2 56 33.09 &  $-$0 03 57.5 & 2.381  &   19.21  &   $+$0.04 & d \\
 2 56 37.05 &  $-$0 34 34.7 & 0.361  &   17.41  &   $-$0.01 & d \\
 2 56 47.40 &  $+$1 46 56.6 & 1.016  &   19.15  &   $-$0.01 & d \\
 2 56 48.18 &  $+$0 46 35.0 & 1.853  &   18.78  &   $+$0.02 & d \\
 2 56 55.14 &  $-$0 31 54.0 & 1.998  &   17.12  &   $-$0.01 & d \\
 2 57 02.25 &  $-$0 20 56.4 & 1.298  &   19.08  &   $+$0.07 & d \\
 2 57 03.26 &  $+$0 25 42.7 & 0.535  &   16.25  &   $+$0.02 & d \\
 2 57 23.70 &  $+$0 23 01.6 & 0.820  &   18.98  &   $-$0.02 & d \\
 2 57 43.13 &  $+$1 16 46.2 & 1.356  &   18.06  &   $+$0.07 & d \\
 2 57 50.00 &  $+$1 54 23.9 & 1.085  &   18.57  &   $-$0.01 & d \\
 2 57 54.15 &  $-$1 00 39.7 & 2.006  &   18.78  &   $+$0.08 & d \\
 2 57 56.08 &  $-$0 07 30.0 & 0.761  &   19.41  &   $+$0.11 & d \\
 2 57 59.68 &  $+$0 31 33.7 & 0.806  &   19.14  &   $+$0.01 & d \\
 2 58 02.48 &  $-$0 27 23.6 & 1.435  &   17.99  &   $+$0.06 & d \\
 2 58 07.68 &  $+$0 20 47.2 & 1.112  &   18.52  &   $-$0.03 & d \\
 2 58 10.43 &  $+$2 10 54.7 & 2.521  &   17.48  &   $+$0.00 & a \\
 2 58 11.37 &  $+$0 05 06.6 & 1.727  &   18.63  &   $-$0.01 & d \\
 2 58 11.53 &  $+$0 09 42.8 & 1.497  &   19.35  &   $+$0.23 & d \\
 2 58 14.54 &  $+$0 42 50.7 & 0.661  &   18.46  &   $+$0.06 & d \\
 2 58 14.72 &  $+$1 37 06.2 & 1.302  &   18.74  &   $+$0.04 & d \\
 2 58 25.73 &  $+$1 37 39.1 & 0.595  &   18.63  &   $+$0.00 & d \\
 2 58 54.46 &  $+$1 45 50.4 & 1.349  &   18.87  &   $+$0.04 & d \\
 2 59 02.77 &  $+$1 12 53.7 & 2.316  &   18.74  &   $+$0.00 & d \\
 2 59 03.14 &  $+$1 26 27.3 & 1.578  &   18.38  &   $+$0.06 & d \\
 2 59 06.29 &  $+$1 04 03.8 & 1.770  &   18.48  &   $+$0.01 & d \\
\hline
\multicolumn{6}{c}{ } \\
\multicolumn{6}{l}{References} \\
\multicolumn{6}{c}{ } \\
\multicolumn{6}{l}{a- \cite{cri:lf}} \\
\multicolumn{6}{l}{b- \cite{bar:cri}}\\
\multicolumn{6}{l}{c- \cite{ver:ver}}\\
\multicolumn{6}{l}{d- \cite{lf:cri}} \\
\multicolumn{6}{l}{e- \cite{papI}}\\
\multicolumn{6}{c}{ } \\
\hline
\end{tabular}
\end{flushleft}
\end{table}

\subsection {Calibration of the photographic material and error estimation}

8 plates taken with the ESO La Silla and UK Schmidt telescopes and a copy of a
POSS E-plate (E1453) have been analysed. 
In Table~\ref{tab:plates} a detailed list is given. 
The combination 
IIIa-F + RG630 filter was used in all cases except for the plate E1453, that
is a combination of a IIIa-E + a red Plexiglas 2444, providing a transmission
similar to that of a Wratten $n^{\circ} 29$ filter (peak sensitivity near
$\lambda 6563$).
The natural photometric system defined by the IIIa-F + RG630 combination 
(in the following defined $R'$)
is close to the Johnson-Kron-Cousins R and the color transformation 
is given by Bessel (1986):
\begin{eqnarray}
R' &=& R-0.013-0.201~(R-I)+0.100~(R-I)^2+\nonumber \\ & &+0.0295~(R-I)^3
\label{eq:bess}
\end{eqnarray}
The combination IIIa-E + red Plexiglas 2444 provides a system for our purposes
indistinguishable from the previous one, with no significant color term in the
range $0.5 < B-R < 3$, as verified on our photometric material.

\begin{table}
\caption[ ]{List of the Plates}  
\label{tab:plates}
\begin{flushleft}
\begin{tabular}{cclcc}
\hline
$Plate$ & $Exp$   & $Date$ & $Epoch/date$ & $Limit$\\
        & $(min)$ &        &              & $(mag)$\\
\hline
E1453~  & 40-60& 1955 Oct 22 & 1 (1955.81) & 19.5 \\
R5413~  &   60 & 1983 Dec 08 & 2 (1983.94) & 20.3 \\
R5414~  &   60 & 1983 Dec 08 & 2 (1983.94) & 20.3 \\
R12218  &  150 & 1987 Oct 16 & 3 (1987.79) & 20.1 \\
R12807  &  140 & 1988 Oct 15 & 4 (1988.79) & 19.5 \\
R13409  &  100 & 1989 Oct 28 & 5 (1989.83) & 20.1 \\
R13420  &  100 & 1989 Nov 01 & 5 (1989.83) & 20.1 \\
R8757~  &  120 & 1990 Sep 21 & 6 (1990.72) & 19.1 \\
R9477~  &  120 & 1991 Oct 07 & 7 (1991.77) & 17.3 \\
\hline
\end{tabular}
\end{flushleft}
\end{table}

The plate material has been scanned with the COSMOS microdensitometer
(\cite{mg:sto}) in IAM mode. The resulting tables, one per each plate,
containing the instrumental magnitudes and other useful parameters for the
objects detected, have been merged together in one table. Only objects with at
least 3 detections in the 9 plates have been accepted in this final table
containing 82\,984 entries. The astrometric error box defining a common
detection has been defined as a circle of 3.5 arcsec of radius. In this way
spurious detections (plate defects) are minimized to an acceptable level, while
real measurements are in practice never discarded. 

%``First-guess'' photometric errors have been estimated in the following way:
%the instrumental magnitudes of all the plates have been transformed in a common
%scale and the distribution of the magnitude deviations of the objects have been
%analysed to derive for each plate the photometric uncertainty as a function of
%the instrumental magnitude. 

\begin{figure}[t]
\epsfxsize=88truemm
\epsffile{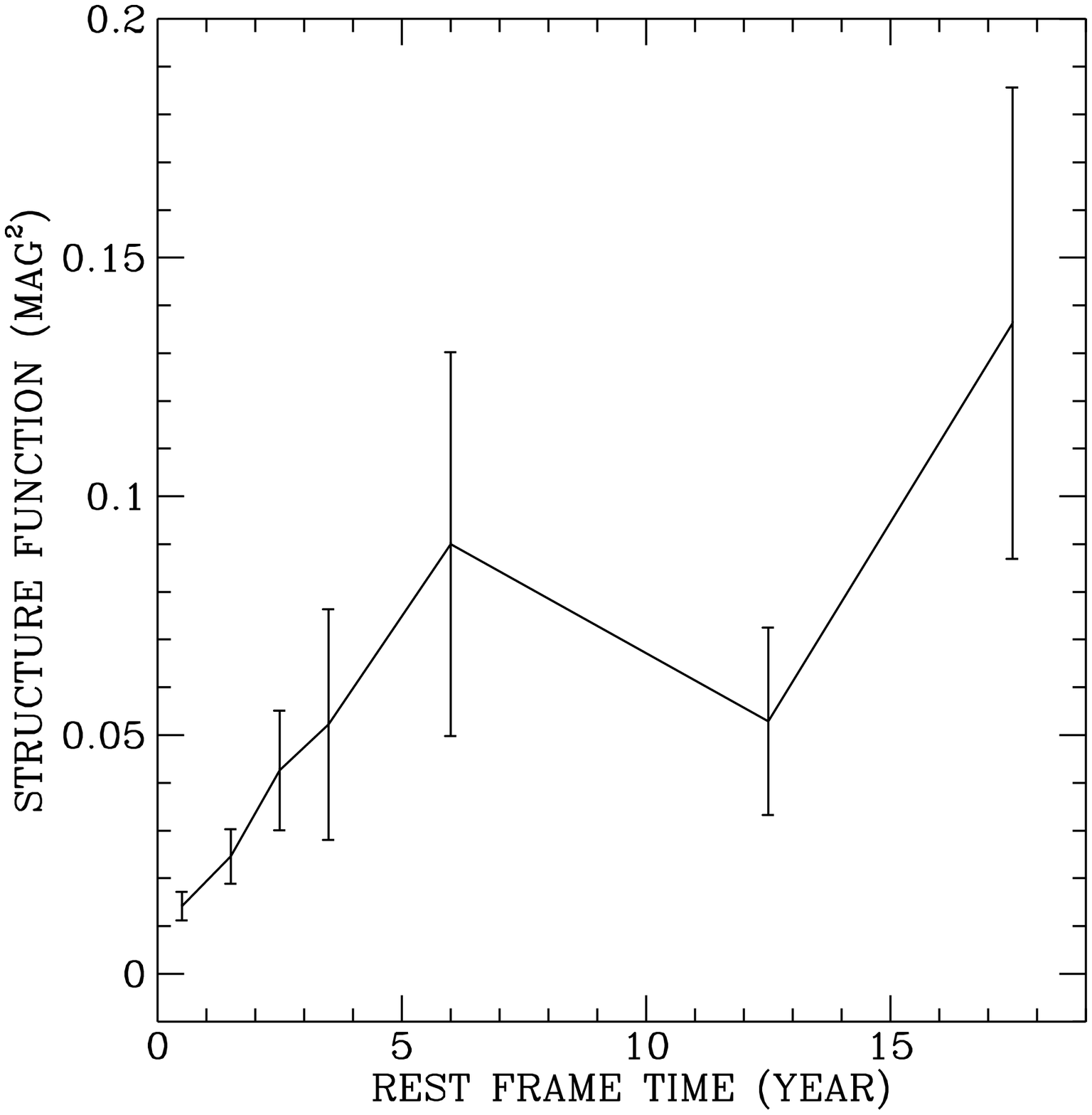}
%\picplace{2cm}
\caption{QSO rest-frame structure function in the R-band}
\label{fig:sfr}
\end{figure}

The calibration of the magnitudes of each plate has been 
carried out according to the following procedure (see also \cite{papI}): 
\begin{enumerate}
\item the magnitudes of 108 standard stars, transformed into the natural
colour system of the plates, $R'$, according to Eq. \ref{eq:bess}, have been
used to derive for each plate a polynomial regression between instrumental
and calibrated magnitudes.
\item the {\it median magnitude} of all the objects has been computed and
each plate has been re-calibrated against the median magnitudes of the
27\,111 point-like objects.
\item a procedure of uniformization of the usually spatially variable
response of the photographic plates has been applied. Each plate has been
subdivided in $10 \times 10$ sub-areas; for each of them the differences
between the reference and the individual plate magnitudes have been computed
and their distribution analysed. 
The zero-point shifts estimated in this way for each sub-area as a function of
the magnitude have been smoothed and applied to the re-calibrated magnitudes. 
In the following we will refer to the magnitudes obtained in this way as
$R'_{final}$.
\item for each plate the uncertainties of the $R'_{final}$ magnitudes have been
estimated by analysing the distribution of the differences $ \Delta R =
R'_{final} - R'_{median}$ as a function of the median magnitudes for all the
point-like objects.
\item the completeness limit of each plate (see Table 2) has been
estimated as the maximum of the histogram of the calibrated magnitudes.
\end{enumerate}

% sono stati scartati gli oggetti con abs_mag>-23 e quelli
% con app_mag>19.8 da cui i 153 oggetti

\begin{figure}[t]
\epsfxsize=88truemm
\epsffile{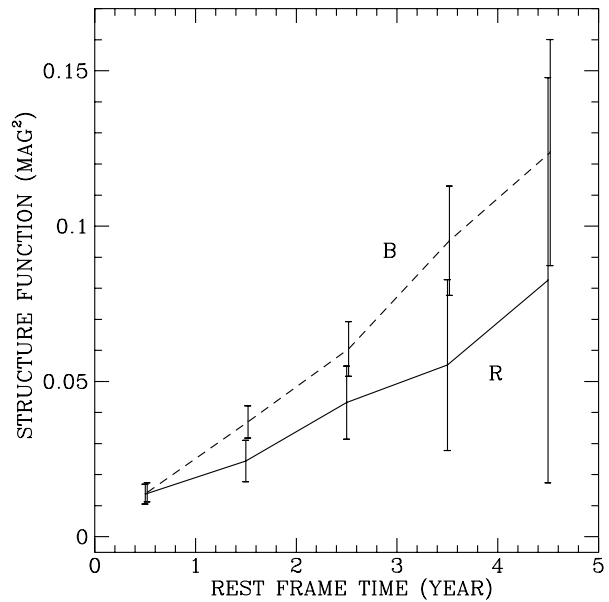}
%\picplace{2cm}
\caption{Structure function for QSOs in the R-band (continuous line) and in
the B-band (dashed line)}
\label{fig:sfbr}
\end{figure}

\section{Statistical indices of the variability}

\subsection {The structure function}

To study the ensemble variability properties of the SA94 sample we have used
the structure function (\cite{sim:cor}, \cite{papI}), defined as:
\begin{equation}
     SF(\tau) = \langle[mag(t)-mag(t+\tau)]^2\rangle      \label{eq:SF}
\end{equation}
where {\em mag(t)} are the magnitudes at the time $t$, $\tau$ is a time lag
evaluated in the rest-frame of each QSO and the brackets `` $\langle \dots
\rangle$ '' indicate a mean over the ensemble. The effect of the measurement
errors has been subtracted according to the procedure described in detail in
\cite{papI}. Only objects detected in at least 5 epochs and 
with a median magnitude brighter than the completeness limit
have been used to compute the SF.
Only 6 QSOs have been measured on Plate R9477 brighter than the completeness
limit. We have checked that the SF and the following statistical results
do not depend on the inclusion/exclusion of these data.

\begin{figure}[t]
\epsfxsize=88truemm
\epsffile{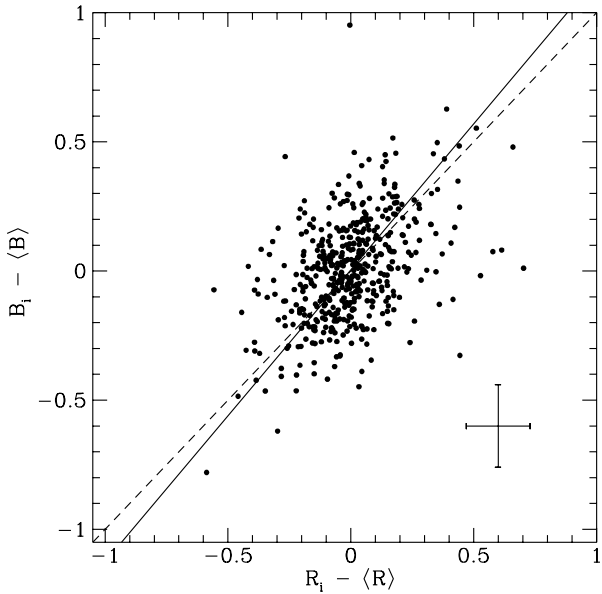}
%\picplace{2cm}
\caption{$ \Delta B_{i} $ vs. $ \Delta R_{i} $ (see text).
	 Dashed line: straight line with a slope 1.
	 Continuous line: least-squares fit between $ \Delta B_{i} $
	 and $ \Delta R_{i} $ (see text). The cross in the lower-right
	 corner shows the typical measurement errors}
\label{fig:deltabr}
\end{figure}

In Fig.~\ref{fig:sfr} the R-band $SF$ in the quasar rest frame is shown.
We observe a steady rise, steeper up to about $ 5 $ years
and a flattening afterwards (within the errors), reaching values
corresponding to about $0.1$ mag$^2$ of RMS variability.

It should be noted that points in Fig.~\ref{fig:sfr} with $\Delta t >
10 yr$ are all derived from the comparison of the POSS plate with the
remaining plates. Being 28 years apart from Plate R5413,
Plate E1453 does not play any role in the estimates
of the SF on the shorter timescales (5 yr or less).

In Fig.~\ref{fig:sfbr} a comparison between the $SF$s in the B-band 
(as derived from the data discussed in Paper I) and in the
R-band is reported. The $SF$ in the B-band shows a larger variability amplitude
with respect to the one in the R-band. 
To quantify the statistical difference between the two $SF$s, we have
calculated the weighted average of the ratio $ SF_{\rm R}/SF_{\rm B} $, for
lags between 1 and 5 yr in the QSOs rest frame. The result is $ SF_{\rm
R}/SF_{\rm B} = 0.66\pm0.13 $, a difference from the unity significant at a
$2.5 \sigma$ level.

In order avoid any possible bias deriving from different time
samplings or time dilation effects, a direct comparison of the
variability in the R and B band has been carried out.
As shown in Fig.~\ref{fig:deltabr}, for each object we have
computed the quantities $ \Delta R_{i} = R_{i} - \langle R \rangle $, as the
difference between the magnitude at a given epoch and the average magnitude. 
Since the epochs of the R plates are in large part coincident with the epochs
sampled with the B plates
(epochs 2, 3, 4, 5, 6, 7 of the present paper correspond to
epochs 2, 6, 8, 9, 10, 11+12 of Paper I, respectively)
it has been possible to plot the $ \Delta B_{i} $
versus the $ \Delta R_{i} $ for all the objects with at least 5
epoch-magnitudes in common.
Again, the POSS plate does not play any role in this comparison.

We have then fitted a straight-line relationship between $ \Delta B_{i} $
and $ \Delta R_{i} $, with a least-squares technique that takes into account
the errors of the data points on both axes (\cite{fas:vio}).
The result is $ \Delta B = (1.130\pm0.054) \Delta R + (0.004\pm0.009) $.
The dispersion of the points around this relation is fully compatible
with the measurement errors, with no intrinsic scatter. If we make
the additional assumption that the R flux varies in phase with the B flux,
i.e. that for $ \Delta R = 0 $ $ \Delta B = 0 $, then we obtain
$ \Delta B = (1.130\pm0.054) \Delta R $,
again fully consistent with no intrinsic scatter. The coefficient between
the R and the B variability obtained in this way is consistent within
the errors with what has been estimated on the basis of the SF analysis.

%We stress here that the form of the ensemble structure function is the result
%of the combination of the intrinsic properties of the QSOs that contribute in
%different proportions to each bin of time (for example the larger $\Delta t$'s
%are populated mainly by lower-redshift objects) and is truly representative
%only if all the QSOs in the samples behave more or less in the same way. This
%is not the case, as will be shown below. 

\begin{figure}[t]
\epsfxsize=88truemm
\epsffile{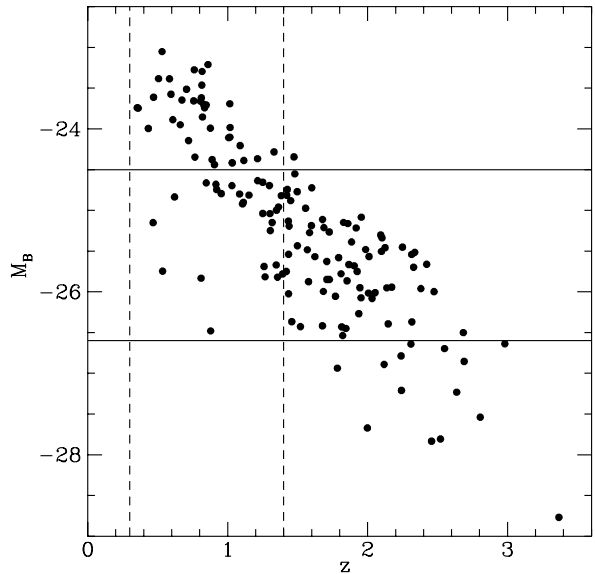}
%\picplace{2cm}
\caption[ ]{Distribution of the SA94 QSO sample in the $M_{\rm B}-z$ plane}
\label{fig:zB}
\end{figure}

\begin{figure}[t]
%\picplace{2cm}
\epsfxsize=88truemm
\epsffile{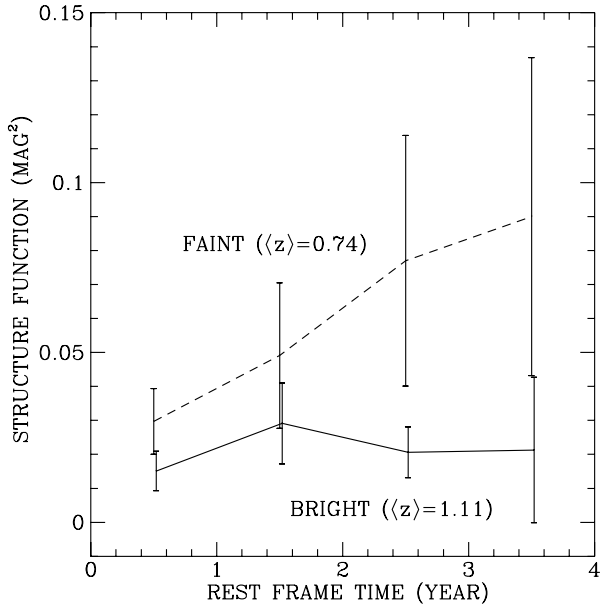}
\caption {The structure function for quasars of different luminosity
	  in the redshift interval $0.3<z<1.4$. Continuous line: QSOs with
	  $M_{\rm B} \leq -24.5$. Dashed line: QSOs with $-24.5 < M_{\rm B}$}
\label{fig:sfM}
\end{figure}

\begin{figure}[t]
\epsfxsize=88truemm
\epsffile{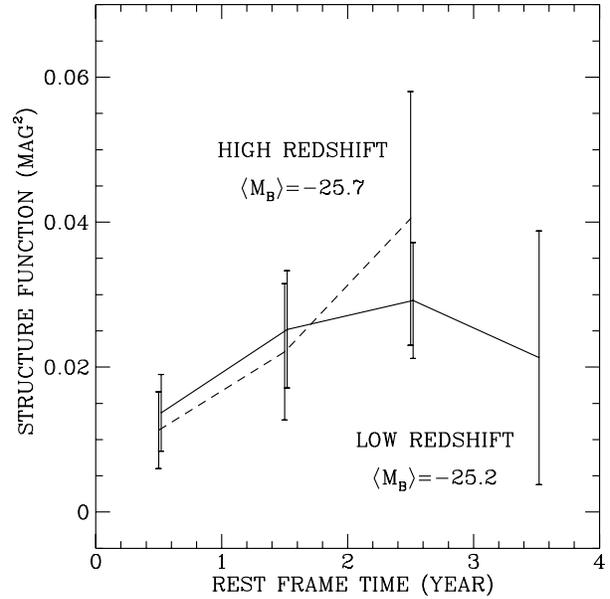}
%\picplace{2cm}
\caption{The structure function for quasars of different redshifts in the
	 luminosity interval $-26.6 < M_{\rm B} < -24.5$. Dashed line: QSOs
	 with $ z < 1.5 $. Continuous line: QSOs with $ 1.5 \leq z $}
\label{fig:sfz}
\end{figure}

In Fig.~\ref{fig:zB} the distribution of QSOs in the redshift-magnitude plane
is illustrated. 
To further investigate and disentangle the dependences of the variability on
the luminosity and redshift we have examined two subsamples.
The first is defined within the redshift limits $0.3 < z < 1.4$ (the area
between the dashed lines in Fig.~\ref{fig:zB}). 
The second is defined within the absolute magnitude limits
$-26.85<M_{\rm B}<-24.50$ (the area between the continuous lines in 
Fig.~\ref{fig:zB}).
As in \cite{papI}, the $SF$s indicate that the QSO variability (in magnitude) 
is, also in the R band, 
clearly anticorrelated with absolute luminosity, in the sense that more
luminous objects have a smaller variability amplitude (Fig.~\ref{fig:sfM}).
No correlation between redshift and
variability - of the type discovered in Paper I for the B-band variability -
has been detected within the errors (Fig.~\ref{fig:sfz}). 
However, such a failure is not surprising considering that in the present work
we are dealing with only one sample (in place of the three of Paper I) and the
still non-uniform coverage of the redshift-absolute magnitude plane of the
subsample defined between $-26.85<M_{\rm B}<-24.50$ introduces a
luminosity-redshift correlation in the data that may mask any
redshift-variability correlation. 

\subsection {The variability index}

In order to further verify the results obtained in the previous section, we have
evaluated the variability index ($IDX$) for each QSO, defined as 
the variance of the normal process representing the intrinsic variability
that, added to the photometric errors, reproduces the observed light variations
on a given range of timescales (see \cite{papI}). 
The index has been evaluated in one time bin 
(from 1 to 15 years in the QSOs rest frame) 
to avoid on the one hand time scales for which the structure function is
still quickly rising, as observed in the previous section, and on the other
hand time lags unaccessible with the present data for the higher-redshift QSOs.
Then we have analysed the correlation matrix, reported in Table~\ref{tab:cc1}.

\begin{table}
\caption[ ]{Correlation coefficients}  \label{tab:cc1}
\begin{flushleft}
\begin{tabular}{cccc}
\hline
 & $R$ & $z$ & $M_{\rm B}$ \\
\hline
$IDX$   &  $0.09\pm0.08$ & $-0.18\pm0.08$ &  $0.24\pm0.08$ \\
$M_{\rm B}$ &  $0.42\pm0.07$ & $-0.81\pm0.03$ &       $-$      \\
$z$     &  $0.15\pm0.08$ &       $-$      &       $-$      \\
\hline
\end{tabular}
\end{flushleft}
\end{table}

The correlation coefficients indicate a correlation of the variability index
with the absolute magnitude and an almost equal anticorrelation with redshift
(correlation coefficients respectively of $+0.24$ and $-0.18$, both with a
significance greater than 99\%). 
Due to the strong flux-redshift anticorrelation in the QSO sample it is not
possible to disentangle if one of the two correlations is spurious. 
However, if, on the basis of the results of the previous subsections and of
\cite{papI}, we assume as fundamental the absolute magnitude - variability
correlation, it is possible to subtract its influence from the
variability-redshift anticorrelation via the method of partial correlation
analysis (\cite{spi}). 
Applying this recipe to the results of Table~\ref{tab:cc1}, 
the corrected correlation coefficient between
variability index and redshift, turns out to be 
$0.03\pm0.08$ (65\%), not significant, in agreement with the
result shown in Fig.~\ref{fig:sfz}.

\section {Discussion}

The decrease of the variability amplitude with increasing wavelength
observed in 
the present QSO sample is consistent with the results obtained 
for individual objects at low
redshift by Edelson et al. (\cite{ede:kro}), Kinney et al. (\cite{kin:91}),
Paltani \& Courvoisier (\cite{pal:cou}).
Is this variability-frequency correlation sufficient to account for 
the positive
correlation between redshift and variability found in \cite{papI} ?

Let us assume a simple power-law representation of a typical QSO
spectrum (\cite{gtv:91}, \cite{dicl:96}):
\begin{equation}
F_{\lambda} \propto \left( \frac {\lambda_0}{\lambda} \right) ^{\alpha}
\end{equation}
where $ F_{\lambda} $ is the flux at an observed wavelength $\lambda$,
$\lambda_{0}$ an arbitrary wavelength and $\alpha$ the spectral index.
The variability can then be modeled as a change in the spectral index:
\begin{equation}
F_{\lambda}(t) \propto 
	       \left(\frac{\lambda_0}{\lambda}\right)^{\alpha+\epsilon(t)}
\end{equation}
where $\epsilon(t)$ is a function of time. In this way the flux variations
are idealized as changes in the slope of the power-law ``hinged''
on the point of wavelength $\lambda_0$, that we imagine somewhere redwards
of the B and R bands. 

\noindent Converting in magnitudes, 
\begin{equation}
M_{\rm R}(t) = 2.5\log\left[\frac{\lambda_{\rm R}}
		{\lambda_{0}(1+z)}\right]^{\alpha+\epsilon(t)} + C ,
\end{equation}
we obtain the following expression for the magnitude variations:
\begin{eqnarray}
\label{eq:DM}
\Delta M_{\rm R} &=& M_{\rm R}(t_2)-M_{\rm R}(t_1)= \nonumber \\
&=& 2.5 [ \epsilon(t_1) - \epsilon(t_2)]
\log\left[ { {\lambda_0 (1+z)} \over \lambda_{\rm R} }\right]
\end{eqnarray}

\noindent Applying the same equation to the B-band, we can write:
\begin{equation}
\label{eq:ratio}
\frac{\Delta M_{\rm B}}{\Delta M_{\rm R}} =
\frac{\log\frac{\lambda_{0}}{\lambda_{\rm B}}+\log\left(1+z\right)}
     {\log\frac{\lambda_{0}}{\lambda_{\rm R}}+\log\left(1+z\right)}
\end{equation}

From Eq.~\ref{eq:ratio} we would expect a decreasing ratio
${\Delta M_{\rm B}}/{\Delta M_{\rm R}}$ with increasing redshift.
In the present data we observe, if any, a very small decrement
of ${\Delta M_{\rm B}}/{\Delta M_{\rm R}}$ with the redshift (from
$1.14\pm0.08$ to $1.10\pm0.09$ for two sub-samples with redshift
$\rm z < 1$ and $1 < \rm z \le 1.8$ respectively).

Introducing the value of ${\Delta M_{\rm B}}/{\Delta M_{\rm R}}=1.13\pm0.05$
at a $\langle z \rangle \sim 1.5$
 reported in section 3 in Eq.~\ref{eq:ratio}, 
allows us to put a lower limit to the
wavelength $\lambda_0 \magcir 31\,000$ \AA. 

If we turn back to Eq.~\ref{eq:DM}, we can see how this lower limit to
$\lambda_0$, together with the typical observed RMS for the variability
${\Delta M_{\rm B}}$ and ${\Delta M_{\rm R}}$ of 0.35 and 0.28 mag,
respectively (see Fig.~\ref{fig:sfbr}), 
puts an upper limit to the RMS variation of the spectral index
$\sigma_{\epsilon} \mincir 0.11$.

This value can be compared with the typical dispersion in the spectral indices
observed in various QSO samples: 
\cite{pei:91}  have obtained $\sigma_\alpha \sim 0.24$ in a control
sample of 15 QSOs, 
Francis (1993) observed $\sigma_\alpha \sim 0.5$ in a much larger sample, 
showing also
a systematic hardening of the spectrum with increasing redshift.
In this way the dispersion in the observed spectral indices of the QSO
population can be only partially accounted for by variability and 
significant intrinsic differences between QSO and QSO have to be
invoked
(but see also the dispersion $\sigma_{\alpha_{ox}} \sim 0.1$ 
estimated by Elvis et al., 1994).

Putting in Eq.~\ref{eq:DM} our best estimates for $\sigma_\epsilon = 0.07$
and $\lambda_0 = 86\,000$ \AA, we obtain for the typical RMS
variability in the $B$ band $0.26$ and $0.33$ mag at $z=0.5$ and $z=3$,
respectively and $0.22$ and $0.30$ in the $R$ band at the same redshifts,
consistent with the observations of Paper I.

The dependence of the variability time scale on the wavelength is a standard
prediction of the accretion disk model. The temperature of the disk decreases
with the radius, and the dynamical, thermal and viscous time scales all
increase as $r^{3/2}$ (see \cite{bag:mal} and references therein). In
particular, in the standard model, involving instabilities in a thin
accretion disk around a supermassive black hole, the intrinsic
characteristic time scale should vary as $\tau \propto \lambda^{2}$.
More in general, if the source is thermal and if the flux variability is due
to temperature changes, the observed wavelength dependence can be naturally
produced (Paltani \& Courvoisier \cite{pal:cou}).
The only requirement concerns the turnover in the spectral energy
distribution that should not be located at such a high frequency that the
spectral slope in the observed band becomes independent of the temperature
(e.g. the Rayleigh-Jeans domain for a black body spectrum).
As remarked by Di Clemente et al. (1996), if brighter objects are on 
average hotter and the variability is due to temperature changes, one
would expect a negative correlation between amplitude of the
variability and luminosity, as observed.
Assuming for example a spectrum of black body or thermal bremsstrahlung,
the spectral turnover of brighter objects is progressively shifted at
higher frequencies producing progressively smaller flux changes, from
$\delta I/I \propto \delta T/T(h\nu/kT)$ for $kT<h\nu$, to
$\delta I/I \propto \delta T/T$ for $kT>h\nu$.

The broad-band variability in the starburst model (\cite{itziar:96}
and references 
therein) is defined by the superposition of a variable component, supernova
explosions (SNe) generating rapidly evolving compact supernova remnants (cSNR),
and a non-variable component, a young stellar cluster and the other stars of
the galaxy. 
The problem of predicting the wavelength dependence of the variability can be
reduced to computing the spectra of the variable and non-variable components,
and their relative luminosities. 
The spectrum of the non-variable component has been predicted to show a
$F_{\nu} \propto \nu^\alpha$ with $\alpha \sim -1.0 \div 0$ (\cite{cid:95}).
The variable/non-variable relative luminosities can also be estimated on the
basis of stellar evolution (\cite{itziar:94}). Much more difficult is to
predict the optical/UV spectrum of the variable component. 
It is expected to be harder than the non-variable component with a 
$F_{\nu} \propto \nu^\alpha$ and $\alpha \sim 0.5$.
We can model the total QSO flux according to a Simple Poissonian model
(\cite{cid:96}) as 
\begin{eqnarray}
\label{eq:stburst}
F_{\rm tot} &=& F_{\rm var} + F_{\rm bck}= \nonumber \\ &=& (1 - f_{\rm bck})
\left( \frac{\lambda}{\lambda_{0}}\right)^{(\alpha_{\rm var}-2)} +
f_{\rm bck}\left(\frac{\lambda}{\lambda_{0}}\right)^{(\alpha_{\rm bck}-2)}
\end{eqnarray}
where $f_{\rm bck}$ is the non-variable fraction of the total flux and
$\alpha_{\rm bck}$, $\alpha_{\rm var}$ are the slopes of the non-variable and
variable components.
Indicating with $\lambda_{\rm B}$ and $\lambda_{\rm R}$ the effective 
wavelengths of the B and R-band and
putting for simplicity $\lambda_{0} = \lambda_{\rm B}$,
we obtain for the ratio between the B-band and R-band variability:
\begin{equation}
\label{eq:ratiosb}
\frac{\Delta M_{\rm B}}{\Delta M_{\rm R}} =
\frac { f_{\rm bck} \left(\frac {\lambda_{\rm R}}{\lambda_{\rm B}}\right)
^{\Delta \alpha} + f_{\rm var}(1+z)^{\Delta \alpha}}
{f_{\rm bck} + f_{\rm var}(1+z)^{\Delta \alpha}}
\end{equation}
with $\Delta \alpha = \alpha_{\rm bck} - \alpha_{\rm var}$. \\
For example, with $\Delta \alpha = 1$ and $f_{\rm bck}, f_{\rm var} = 
0.7, ~ 0.3$
respectively (\cite{cid:96}), the ratio between the B-band and R-band 
variability is expected to be about 1.4 at $z=0$, tending to 1
for increasing redshift, 
independent on the intrinsic luminosity of the QSO. 
The above reported slight tendency
for a decreasing ${\Delta M_{\rm B}}/{\Delta M_{\rm R}}$ ratio
with increasing redshift is compatible with this model.
However,
an increase of the ${\Delta M_{\rm B}}/{\Delta M_{\rm R}}$
with the absolute luminosity is also observed.
We find the values ${\Delta M_{\rm B}}/{\Delta M_{\rm R}}=1.08\pm0.06$ and
${\Delta M_{\rm B}}/{\Delta M_{\rm R}}=1.31\pm0.11$ for two subsamples with
$M_{\rm B} > -25.2$ ($\langle z \rangle = 1.05$) and 
$M_{\rm B} < -25.2$ ($\langle z \rangle = 1.92$), respectively. 
This suggests that more general situations 
have to be envisaged
in which the
fraction on the non-variable component and/or the pulse properties
(e.g. the spectral distribution) depend on the global 
luminosity of the object.

Recently Hawkins (\cite{hawk:96}) has proposed that the variability of nearly
all the QSOs (except the very low redshift ones) is produced by microlensing.
A non-achromatic behaviour can be accommodated in this kind of model. For a
source comparable in size to the Einstein ring of the lens, with a quasar
disk redder at larger radii, the bluer compact core would produce a larger
amplitude of the variations 
in the blue passband whereas the larger extent of the red image would
cause a small amplitude variation, albeit with a longer duration, in the red.
However it cannot be denied that achromatic light variations would have been
one of the most significant characteristics 
of the microlensing model and the present
observation of the wavelength dependence make it less appealing/compelling.
In particular the superposition of the two competing effects of time
dilation and correlation of the variability with frequency, i.e. redshift,
accounts naturally, also in the accretion disk and in the starburst scenarios,
for the observed lack of correlation in the observer's rest 
frame between the variability time scale and the redshift, otherwise
claimed as an evidence in favor of the microlensing model.

Future detailed comparisons between simulated and observed light curves
will be fundamental to disentangle among these three scenarios and to
start exploring the parameter space of each model.

\acknowledgements
%________________________________________ Do not leave a blank line here!
It is a pleasure to thank I. Aretxaga, G. Fasano, E. Giallongo, 
M. Hawkins, R. Terlevich, D. Trevese, M. H. Ulrich,
P. V\'eron  and R. Vio for enlightening discussions.
This research has been partially supported by the ASI contracts
94-RS-107 and by ANTARES, an astrophysics network funded by the HCM programme
of the European Community.                           
{}
\end{document}